\documentclass[9pt,conference]{IEEEtran}
\IEEEoverridecommandlockouts
\usepackage{cite}
\usepackage{amsmath,amssymb,amsfonts}
\usepackage{algorithmic}
\usepackage{graphicx}
\usepackage{textcomp}
\usepackage{xcolor}
\usepackage{multirow}
\usepackage{bbding}
\usepackage{booktabs}
\def\BibTeX{{\rm B\kern-.05em{\sc i\kern-.025em b}\kern-.08em
    T\kern-.1667em\lower.7ex\hbox{E}\kern-.125emX}}
\begin{document}

\title{Speech Enhancement with Overlapped-Frame Information Fusion and Causal Self-Attention\\
\thanks{$^\star$ Equal contribution \\ \indent $^\dagger$ Corresponding author}
}

% \name{Author Name$^{\star \dagger}$ \qquad Author Name$^{\star}$ \qquad Author Name$^{\dagger}$}

% \address{$^{\star}$ Affiliation Number One \\
% $^{\dagger}$ Affiliation Number Two}

\author{\IEEEauthorblockN{Yuewei Zhang\textsuperscript{$\star$}}
\IEEEauthorblockA{\textit{Department of Electronic Engineering} \\
\textit{Shanghai Jiao Tong University}\\
Shanghai, China \\
yueweizhang@sjtu.edu.cn}
\and
\IEEEauthorblockN{Huanbin Zou\textsuperscript{$\star$}}
\IEEEauthorblockA{
\textit{Xiaohongshu Inc.}\\
Shanghai, China \\
1518854421@alumni.sjtu.edu.cn}
\and
\IEEEauthorblockN{Jie Zhu\textsuperscript{$\dagger$}}
\IEEEauthorblockA{\textit{Department of Electronic Engineering} \\
\textit{Shanghai Jiao Tong University}\\
Shanghai, China \\
zhujie@sjtu.edu.cn}
}

%\ninept
\maketitle

\begin{abstract}
For time-frequency (TF) domain speech enhancement (SE) methods, the overlap-and-add operation in the inverse TF transformation inevitably leads to an algorithmic delay equal to the window size. However, typical causal SE systems fail to utilize the future speech information within this inherent delay, thereby limiting SE performance. In this paper, we propose an overlapped-frame information fusion scheme. At each frame index, we construct several pseudo overlapped-frames, fuse them with the original speech frame, and then send the fused results to the SE model. Additionally, we introduce a causal time-frequency-channel attention (TFCA) block to boost the representation capability of the neural network. This block parallelly processes the intermediate feature maps through self-attention-based operations in the time, frequency, and channel dimensions. Experiments demonstrate the superiority of these improvements, and the proposed SE system outperforms the current advanced methods.
\end{abstract}


\section{Introduction}

Speech enhancement (SE) aims to suppress the noise components in a noisy audio and recover the target clean speech. In the real-world scenarios, speech signals are frequently contaminated by diverse noises. As a result, the utilization of a SE module as a front-end processing module has become prevalent across various speech-related applications, such as speech recognition, speaker verification, and voice activity detection.

In the past few years, deep leaning (DL) techniques have garnered extensive interest from researchers across various fields. Thanks to the powerful representation capabilities of deep neural network (DNN), the DNN-based SE methods \cite{pascual17_interspeech, defossez20_interspeech, tan18_interspeech, hu20g_interspeech, 9414177, 9431717, cao22_interspeech} have also achieved remarkable performance. Common DNN-based SE methods can be divided into two families, namely time domain methods \cite{pascual17_interspeech, defossez20_interspeech} and time-frequency (TF) domain methods \cite{tan18_interspeech, hu20g_interspeech, 9414177, 9431717, cao22_interspeech}. The time domain methods perform direct noise reduction on the waveform in an end-to-end way, while the TF domain methods operate on the noisy TF spectrum to suppress the noise components. Nowadays, the TF domain SE methods have become mainstream.

\begin{figure}[t]
  \centering
  \includegraphics[width=7cm]{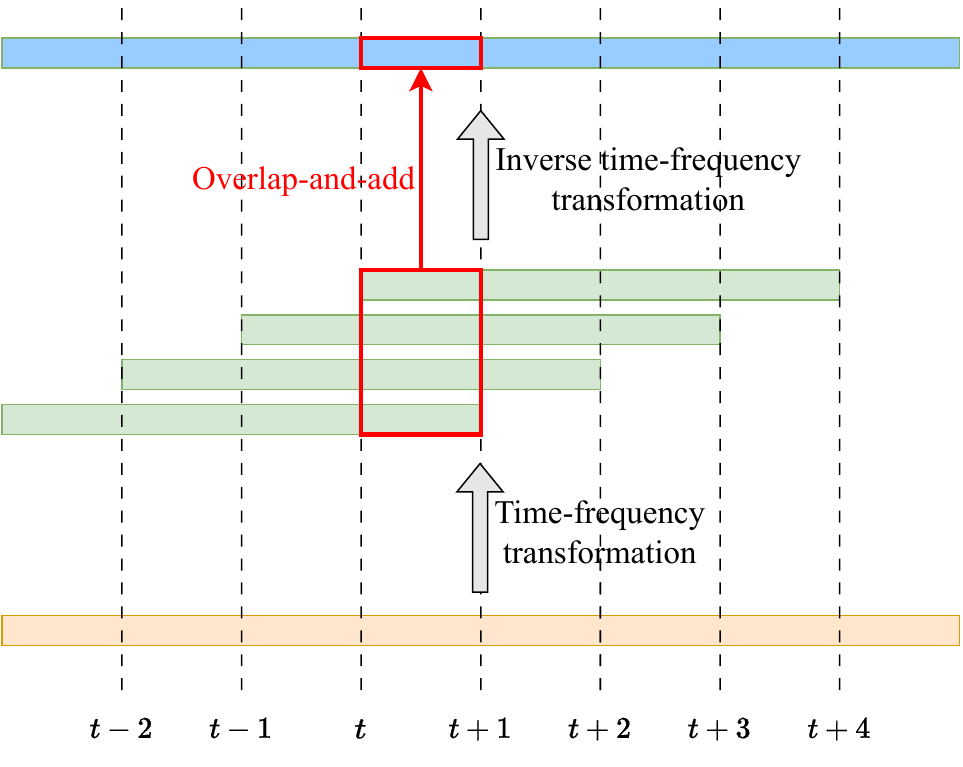}
  \caption{Time-frequency transformation \& Inverse time-frequency transformation.}
  \label{fig1}
\end{figure}

In the practical deployment of SE algorithms, achieving a low algorithmic delay is crucial for ensuring a positive user experience. Therefore, many causal SE methods \cite{defossez20_interspeech, tan18_interspeech, hu20g_interspeech, 9414177, 9431717, 9747888} have been proposed, which utilize only current and past input audio information to generate the current enhanced result. For the TF domain SE methods, since they need to first transform the noisy waveform into the spectrum by a TF transformation, the causal inference implies that the DNN does not look-ahead any feature speech frames when enhancing the current speech frame. After predicting the enhanced spectrum, the inverse TF transformation is employed to reconstruct the denoised waveform. However, as pointed out by \cite{9797050}, there is an inherent algorithmic delay caused by the inverse TF transformation. As shown in Fig.~\ref{fig1}, during the inverse TF transformation process, each reconstructed waveform frame is actually obtained by overlapping and adding multiple consecutive adjacent frames, resulting in an inevitable algorithmic delay equal to the window size. To fully utilize the future information in this inherent algorithmic delay, \cite{9797050} proposes an overlapped-frame prediction scheme. Nevertheless, designing an engineered synthesis window corresponding to the analysis window for perfect reconstruction is challenging and inflexible for different applications and conditions.

\begin{figure*}[t]
  \centering
  \includegraphics[width=17.3cm]{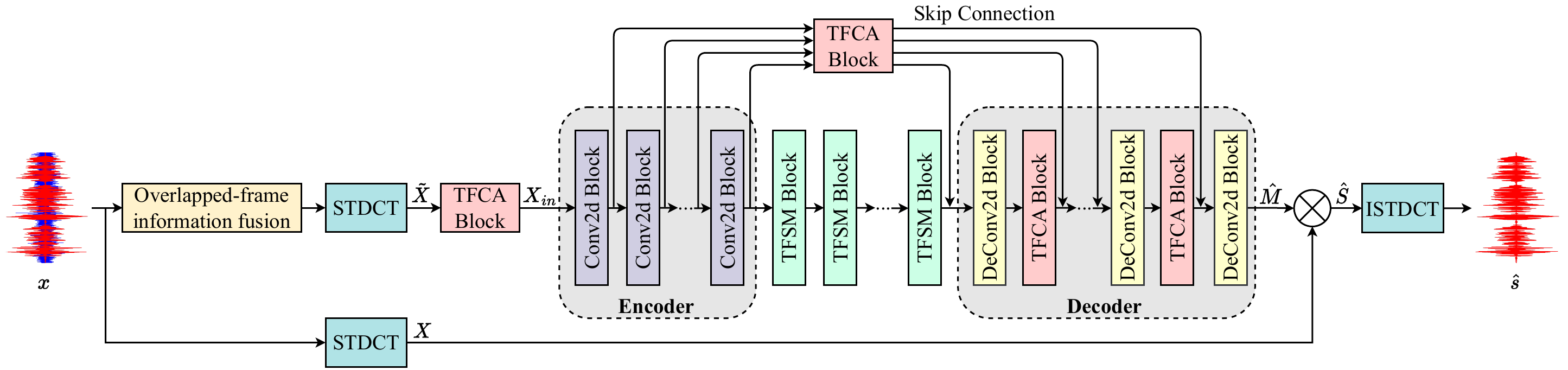}
  \caption{Overall architecture of the proposed SE system combining the overlapped-frame information fusion scheme with self-attention-based DCTCRN (OFIF-Net).}
  \label{fig2}
\end{figure*}

In this paper, we propose an overlapped-frame information fusion (OFIF) scheme. At each frame index, we fuse the real current speech frame with several pseudo future frames, to obtain a fused spectral feature. The utilization of pseudo speech frames can simulate part of the information of future speech frames in the inherent algorithmic delay, thereby achieving more complete information utilization. The fused feature is subsequently processed by a feature post-processing module and fed into the SE DNN for denoising processing.

In addition, the convolutional recurrent network (CRN)-based architecture \cite{tan18_interspeech,hu20g_interspeech,li2021real,9747578,10096918} has been demonstrated to be effective for SE. Therefore, our proposed SE system takes the DCTCRN \cite{li2021real} as the network backbone. To enhance the sequence modeling capability, we adopt the time-frequency sequence modeling (TFSM) block \cite{10447096} to construct the recurrent module in DCTCRN. Furthermore, we devise a time-frequency-channel attention (TFCA) block based on the self-attention mechanism. The TFCA block is inserted into the skip connection path and decoder of DCTCRN. Note that all the neural network parts, including the CRN strcuture, the TFSM block, and the TFCA block, are designed to be causal.

Combining the overlapped-frame information fusion scheme with self-attention-based DCTCRN network, we name the proposed SE system OFIF-Net. Experimental results confirm the benefits of the aforementioned improvements, and OFIF-Net eventually achieves superior performance compared to existing competitive benchmarks.

\section{Methodology}

The overall architecture of OFIF-Net is illustrated in Fig.~\ref{fig2}. The OFIF-Net takes the noisy signal $x=s+z$ as input, where $s$ and $z$ denote the clean and noise signals. It predicts an enhanced audio $\hat{s}$. The main parts of OFIF-Net includes the proposed OFIF block and a SE network, whose details will be introduced as follows.

\subsection{Overlapped-frame information fusion (OFIF)}

Before the operations of SE network, the noisy speech $x$ need to be first transformed to the TF spectrum $X\in\mathbb{R}^{F_{i}\times{T_{i}}}$ by short-time discrete cosine transform (STDCT) \cite{1672377}, where $F_{i}$ and $T_{i}$ denote the number of frequency bins and frames. During the process of STDCT, the speech waveform $x$ need to be first divided into the overlapped short-time speech frames, which are then converted to the frequency domain representations by discrete cosine transform (DCT). In this work, we set the DCT points number $N$, Hamming frame size $W$, and frame shift $H$ as $N=512$, $W=32\ \text{ms}$, and $H=8\ \text{ms}$, respectively. Therefore, there is an overlap among $W/H=4$ adjacent speech frames. Moreover, at each frame index $t$ $(t\in[1,T])$, frame overlap means that the current speech frame $x_t\in\mathbb{R}^{W}$ contains a part of the information of the next 3 speech frames, i.e., $x_{t+1},x_{t+2},x_{t+3}\in\mathbb{R}^{W}$. Consequently, we propose to construct 3 pseudo speech frames by zero-masking $x_t$ as:
\begin{align}
  \Tilde{x}_{t+1} = Mask\{x_{t}\}^{\frac{3}{4}H \sim H}
  \label{equation2}
\end{align}
\begin{align}
  \Tilde{x}_{t+2} = Mask\{x_{t}\}^{\frac{1}{2}H \sim H}
  \label{equation3}
\end{align}
\begin{align}
  \Tilde{x}_{t+3} = Mask\{x_{t}\}^{\frac{1}{4}H \sim H}
  \label{equation4}
\end{align}
where $\Tilde{x}_{t+1},\Tilde{x}_{t+2},\Tilde{x}_{t+3}\in\mathbb{R}^{W}$ are the pseudo speech frames. $Mask\{x_{t}\}^{\frac{3}{4}H \sim H}$ means that the $\frac{3}{4}H \sim H$ fragment of $\tilde{x}_{t+1}$ is masked to zero, while the $0 \sim \frac{3}{4}H$ fragment of $\tilde{x}_{t+1}$ corresponds to the $\frac{1}{4}H \sim H$ fragment in $x_t$. Therefore, the superscript of $Mask\{\cdot\}$ indicates the range of the fragment that is masked to zero. And the masking ways in \eqref{equation3} and \eqref{equation4} are the same.

Subsequently, we apply STDCT to the 4 speech frames $\{x_t,\Tilde{x}_{t+1},\Tilde{x}_{t+2},\Tilde{x}_{t+3}\}$ to obtain $\Tilde{X}_{t}\in\mathbb{R}^{4\times{F_{i}}}$. $\Tilde{X}_{t}$ denotes a spectral feature containing the information of future pseudo speech frames. Using the above method for the speech frames at all frame indexes, we obtain the spectrum $\Tilde{X}\in\mathbb{R}^{4\times{F_{i}}\times{T_{i}}}$ that incorporates information of overlapped future pseudo speech frames. The $\Tilde{X}$ is then fed into a TFCA block, which will be described in the next section, to derive the final input feature $X_{in}\in\mathbb{R}^{4\times{F_{i}}\times{T_{i}}}$ for the SE network. By this way, we make full use of current speech frame and pesudo future speech information without adding additional algorithmic delay.

\subsection{Network structure}

\begin{figure*}[t]
  \centering
  \includegraphics[width=14.5cm]{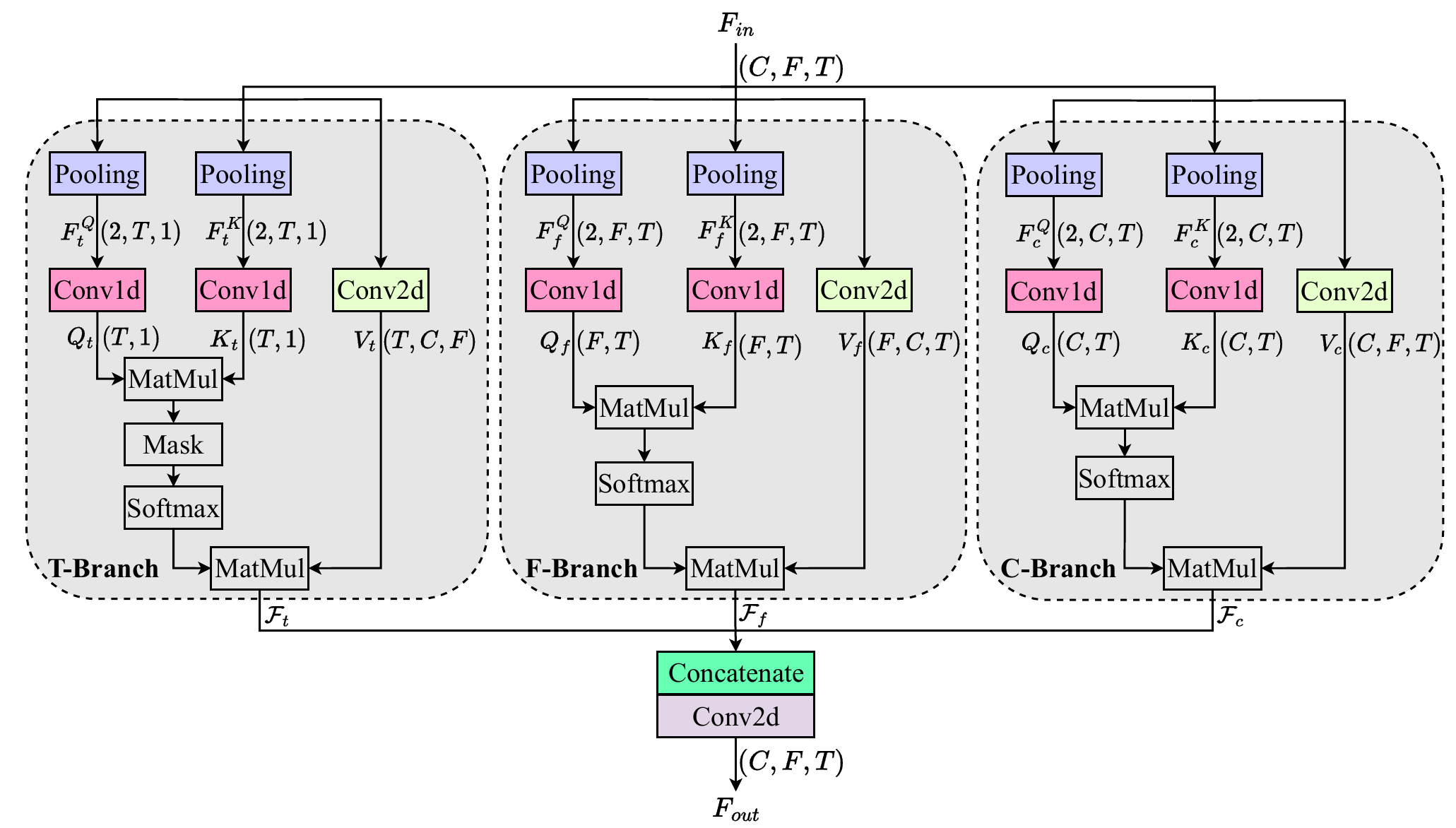}
  \caption{Details of causal time-frequency-channel attention (TFCA) block. The ``T-Branch'', ``F-Branch'', and ``C-Branch'' respectively denote the time-wise, frequency-wise, and channel-wise self-attention branches.}
  \label{fig3}
\end{figure*}

As depicted in Fig.~\ref{fig2}, OFIF-Net adopts a CRN \cite{tan18_interspeech} structure to process $X_{in}$ and perform noise suppression. The CRN structure follows the encoder-decoder (ED) framework, with multiple TFSM blocks \cite{10447096} inserted between the encoder and decoder for sequence modeling. Additionally, the proposed TFCA block is added to both the skip connection paths and decoder layers. The TFCA block boosts the representation capability of the CRN, thereby enhancing the SE performance. Finally, the decoder predict a spectrum mask $\hat{M}$, which is multiplied with the noisy spectrum $X$ to obtain the enhanced spectrum $\hat{S}$ as:
\begin{align}
  \hat{S} = \hat{M}\otimes{{X}}
  \label{equation5}
\end{align}
where $\otimes$ denotes the element-wise multiplication.

\subsubsection{Convolutional recurrent network (CRN)-based structure}

As illustrated in Fig.~\ref{fig2}, the encoder of CRN receives the input feature $X_{in}$ to extract the high-level TF representation. It consists of several 2D convolutional (Conv2d) blocks, each comprising a 2D convolutional layer, a batch normalization, and a PReLU function. The encoded feature is subsequently fed into the TFSM blocks. Similar to the dual-path sequence modeling method in \cite{9054266}, the TFSM block respectively employs a bidirectional gated recurrent unit (BiGRU) layer and a unidirectional gated recurrent unit (GRU) layer to capture sequential correlations along the frequency and time dimensions. Afterwards, the output from the TFSM blocks is sent to the decoder. It aims to reconstruct a spectrum mask $\hat{M}$ with the same TF resolution as the noisy spectrum $X$. Symmetrical to the encoder, the decoder includes several 2D deconvolutional (DeConv2d) blocks. Each DeConv2d block comprises a 2D deconvolutional layer, a batch normalization, and a PReLU function, except that the last DeConv2d block utilizes the Tanh function to ensure a more reasonable range for the spectrum mask estimation.

\subsubsection{Causal Time-frequency-Channel attention (TFCA) block}

To adaptively recalibrate the TF feature maps and further enhance the representation capability of CRN, we devise a TFCA block. The details of TFCA block are shown in Fig.~\ref{fig3}. The TFCA block takes the intermediate feature map $F_{in}\in\mathbb{R}^{C\times{F}\times{T}}$ as input, where $C$, $F$, and $T$ represent the number of channels, frequency bins, and frames. This block generally consists of three parallel branches to capture the global dependencies along the time (``T-Branch''), frequency (``F-Branch''), and channel (``C-Branch'') dimensions, respectively. Meanwhile, we ensure the three self-attention-based branches to be causal for a low algorithmic delay. 

Specifically, in the ``T-Branch'', two pooled results $F_{t}^{Q}\in\mathbb{R}^{2\times{T}\times{1}}$ and $F_{t}^{K}\in\mathbb{R}^{2\times{T}\times{1}}$ are derived by applying global pooling operations to $F_{in}$ along both the channel and frequency dimensions. The resulted channel number denotes two different pooling methods, i.e., average pooling and maximum pooling. The $F_{t}^{Q}$ and $F_{t}^{K}$ are respectively fed into two 1D point-wise convolutional layers to obtain the query and key matrices $Q_{t}\in\mathbb{R}^{T\times{1}}$ and $K_{t}\in\mathbb{R}^{T\times{1}}$ (the channel dimension is omitted). Therefore, the attention matrix $Atten_t\in\mathbb{R}^{T\times{T}}$ is defined as:
\begin{align}
  Atten_{t} = \mathrm{Softmax}(\mathcal{K}({Q_{t}K_{t}^{\top}}))
  \label{equation6}
\end{align}
where $\mathcal{K}\in\mathbb{R}^{T\times{T}}$ represents the temporal causal mask, which is a lower triangular matrix.

In the ``F-Branch'' and ``C-Branch'', the input feature $F_{in}$ is also pooled at first. However, global pooling along the time dimension will lead to non-causal inference. Thus, we adopt local adaptive pooling operation. Take the ``F-Branch'' as an example, we pad $K_{T}-1$ zeros at the beginning of the time dimension of $F_{in}$ and subsequently apply adaptive average pooling with a kernel size of ${K_T}\times{C}$ along the time and channel dimensions. This yields an average pooled result with a shape of $F\times{T}$. Similar methods are employed to achieve causal maximum pooling for $F_{in}$. By concatenating the results of adaptive average pooling and adaptive maximum pooling, we obtain causally pooled results $F_{f}^{Q}\in\mathbb{R}^{2\times{F}\times{T}}$ and $F_{f}^{K}\in\mathbb{R}^{2\times{F}\times{T}}$. Next, $F_{f}^{Q}$ and $F_{f}^{K}$ are respectively processed by two 1D point-wise convolutional layers to derive the query and key matrices $Q_{f}\in\mathbb{R}^{F\times{T}}$ and $K_{f}\in\mathbb{R}^{F\times{T}}$. For the ``C-Branch'', the query and key matrices $Q_{c}\in\mathbb{R}^{C\times{T}}$ and $K_{c}\in\mathbb{R}^{C\times{T}}$ are calculated by a similar method as ``F-Branch''. Afterwards, the attention matrices of these two branches are defined as:
\begin{align}
  Atten_{f} = \mathrm{Softmax}({Q_{f}K_{f}^{\top}}/\sqrt{T})
  \label{equation7}
\end{align}
%\vspace{-\baselineskip}
%\vspace{-\baselineskip}
\begin{align}
  Atten_{c} = \mathrm{Softmax}({Q_{c}K_{c}^{\top}}/\sqrt{T})
  \label{equation8}
\end{align}
where $Atten_f\in\mathbb{R}^{F\times{F}}$, $Atten_c\in\mathbb{R}^{C\times{C}}$. 

Consequently, the outputs of the three branches are:
\begin{align}
    \begin{cases}
    \mathcal{F}_t = Atten_tV_t \\
    \mathcal{F}_f = Atten_fV_f \\
    \mathcal{F}_c = Atten_cV_c \\
    \end{cases}
  \label{equation9}
\end{align}
where $V_t$, $V_f$, and $V_c$ are obtained by processing the input feature $F_{in}$ through three 2D convolutional layers. Finally, the three outputs $\mathcal{F}_t, \mathcal{F}_f,\mathcal{F}_c\in\mathbb{R}^{C\times{F}\times{T}}$ are concatenated and fed into a 2D convolutional layer to generate the TFCA block output $F_{out}\in\mathbb{R}^{C\times{F}\times{T}}$.

\subsection{Loss function}
The loss function $\mathcal{L}$ used for model training is defined as:
\begin{align}
  \mathcal{L} = \lVert{\hat{s}-s}\rVert_1 + \lVert{\hat{M}-M}\rVert_F^2
  \label{equation10}
\end{align}
where $\hat{s}$ and $s$ are the denoised and clean audios, $\hat{M}$ and $M$ are the estimated and target spectrum masks. $\lVert\cdot\rVert_1$ and $\lVert\cdot\rVert_F^2$ denote the L1 loss and mean square error (MSE) loss, respectively.

\section{Experiments}

\subsection{Datasets}

The OFIF-Net is trained on VoiceBank+DEMAND \cite{valentini2016investigating} and DNS-Challenge \cite{reddy20_interspeech} datasets. The VoiceBank+DEMAND dataset comprises 11,572 audio clips for training and 824 audio clips for testing. The training set is generated by mixing clean speech from 28 speakers with 10 types of noise signals at the signal-to-noise ratios (SNRs) of \{0,5,10,15\}dB. The test set adopts the clean audios from 2 unseen speakers to mix with 5 other types of noise at the SNRs of \{2.5,7.5,12.5,17.5\}dB. The DNS-Challenge includes over 500 hours of clean audios and over 180 hours of noise audios. We synthesize 3,000 hours of noisy audios for model training, and the SNR ranges from -5dB to 15dB. The officially provided non-blind validation set containing 150 audio clips is used for model evaluation.

\subsection{Experimental setup}

All audios are sampled at 16 kHz. There are 5 Conv2d block in the encoder, and the output channels of the convolutional layers are \{16,32,64,128,128\}. The decoder correspondingly consists of 5 DeConv2d blocks, with the output channels of the deconvolutional layers configured as \{128,64,32,16,1\}. The (de)convolutional kernel size and stride are set to (5,2) and (2,1), respectively. All the (de)convolutional layers are set to be causal by asymmetric zero-padding and frame-discarding. Furthermore, 3 TFSM blocks are inserted between encoder and decoder, in which the hidden sizes of GRU\&BiGRU layers are \{128,64,32\}. Additionally, we set $K_{T}=15$ for TFCA block.

A RMSprop optimizer with an initial learning rate of 0.0002 is used for model training. If the model performance does not improve for 8 consecutive epochs, the learning rate is halved. The batch size and training epoch are set to 16 and 100.

\subsection{Ablation study}

\begin{table}[t]
  \caption{Results of ablation study.}
  \label{table1}
  \centering
  \setlength{\tabcolsep}{1.8mm}{
  \begin{tabular}{lccccc}
    \toprule
        & WB-PESQ & CSIG  & CBAK  & COVL \\
    \hline
    noisy & 1.97 & 3.35 & 2.44 & 2.63 \\
    \hline
    TF-DCTCRN \cite{li2021real, 10447096} & 2.92 & 4.10 & 3.47 & 3.51 \\
    \quad +OFIF & 2.96 & 4.20 & 3.49 & 3.59 \\
    \quad +TFCA & 2.99 & 4.20 & 3.50 & 3.60 \\
    \quad +OFIF\&TFCA & \multirow{2}*{\textbf{3.06}} & \multirow{2}*{\textbf{4.27}} & \multirow{2}*{\textbf{3.51}} & \multirow{2}*{\textbf{3.67}}\\
\quad (i.e., OFIF-Net) & ~ & ~ & ~ & ~ & ~\\
    \bottomrule
  \end{tabular}}
\end{table}

To demonstrate the benefits of the proposed improvements, We conduct ablation studies on the VoiceBank+DEMAND dataset. We adopt four objective metrics to evaluate the model performance, including wide-band perceptual evaluation of speech quality (WB-PESQ) \cite{rec2005p} and three mean opinion score (MOS)-related metrics \cite{4389058} measuring the signal distortion (CSIG), background noise intrusiveness (CBAK), and overall audio quality (COVL), respectively.

The results of ablation study are illustrated in Table~\ref{table1}. We take DCTCRN \cite{li2021real} combined with the TFSM block \cite{10447096} as baseline model and abbreviate it as TF-DCTCRN. Based on the baseline, we separately verify the benefits of each improvement method. With the help of OFIF scheme, the WB-PESQ, CSIG, CBAK, and COVL are improved by 0.04, 0.10, 0.02, and 0.08, respectively. This demonstrates that constructing pseudo speech frames for more comprehensive information utilization enhances the SE performance. Meanwhile, the incorporation of the TFCA block yields performance gains of 0.07, 0.10, 0.03, and 0.09 on the WB-PESQ, CSIG, CBAK, and COVL metrics, respectively. Therefore, the proposed TFCA block effectively enhances the representation capability of DNN and improves model performance. Eventually, the proposed OFIF-Net is obtained by integrating the aforementioned two improvements. It can be observed that the SE performance is further improved, with the WB-PESQ, CSIG, CBAK, and COVL scores reaching 3.06, 4.27, 3.51, and 3.67, respectively.

\subsection{Comparison with existing advanced methods}

\begin{table}[t]
  \caption{Performance comparison with existing methods on the VoiceBank+DEMAND dataset under causal configuration.}
  \label{table2}
  \centering
  \setlength{\tabcolsep}{1.6mm}{
  \begin{tabular}{lccccc}
    \toprule
        & Para.(M) & WB-PESQ & CSIG  & CBAK  & COVL \\
    \hline
    noisy & - & 1.97 & 3.35 & 2.44 & 2.63 \\
    DCCRN \cite{hu20g_interspeech} & 3.7 & 2.68 & 3.88 & 3.18 & 3.27 \\
    FullSubNet+ \cite{9747888} & 8.67 & 2.88 & 3.86 & 3.42 & 3.57 \\
    CompNet \cite{FAN2023508} & 4.26 & 2.90 & 4.16 & 3.37 & 3.53 \\
    CTS-Net \cite{9431717} & 4.35 & 2.92 & 4.25 & 3.46 & 3.59 \\
    DEMUCS \cite{defossez20_interspeech} & 128 & 2.93 & 4.22 & 3.25 & 3.52 \\
    GaGNet \cite{LI2022108499} & 5.94 & 2.94 & 4.26 & 3.45 & 3.59 \\
    VSANet \cite{10389633} & 3.10 & 2.98 & 4.21 & \textbf{3.51} & 3.60 \\
    \textbf{OFIF-Net} & 2.61 & \textbf{3.06} & \textbf{4.27} & \textbf{3.51} & \textbf{3.67}\\
    \bottomrule
  \end{tabular}}
\end{table}

\begin{table}[t]
  \caption{Performance comparison with existing methods on the DNS-Challenge non-blind test set under causal configuration.}
  \label{table3}
  \centering
  \setlength{\tabcolsep}{1.0mm}{
  \begin{tabular}{lccccc}
    \toprule
        & Para.(M) & WB-PESQ & NB-PESQ  & STOI  & SI-SNR \\
    \hline
    noisy & - & 1.58 & 2.45 & 91.52 & 9.07 \\
    SICRN \cite{10446396} & 2.16 & 2.62 & 3.23 & 95.83 & 16.00 \\
    PoCoNet \cite{isik20_interspeech} & 50 & 2.75 & - & - & - \\
    CTS-Net \cite{9431717} & 4.35 & 2.94 & 3.42 & 96.66 & 17.99 \\
    FullSubNet+ \cite{9747888} & 8.67 & 2.98 & 3.50 & 96.69 & 18.34 \\
    Inter-SubNet \cite{10094858} & 2.29 & 3.00 & 3.50 & 96.61 & 18.05 \\
    FS-CANet \cite{chen22k_interspeech} & 4.21 & 3.02 & 3.51 & 96.74 & 18.08 \\
    GaGNet \cite{LI2022108499} & 5.94 & 3.17 & 3.56 & 97.13 & \textbf{18.91} \\
    \textbf{OFIF-Net} & 2.61 & \textbf{3.30} & \textbf{3.62} & \textbf{97.34} & 18.32 \\
    \bottomrule
  \end{tabular}}
\end{table}

We compare the performance of the proposed OFIF-Net with existing advanced benchmarks, and the results are presented in Table~\ref{table2} and \ref{table3}. When conducting performance comparisons on the DNS-Challenge dataset, WB-PESQ, narrow-band perceptual evaluation of speech quality (NB-PESQ) \cite{941023}, short-time objective intelligibility (STOI) \cite{5713237}, and scale-invariant signal-to-noise ratio (SI-SNR) \cite{8683855} are used as objective metrics. It can be found that our proposed OFIF-Net outperforms previous methods on both datasets. Although GaGNet \cite{LI2022108499} attains a higher SI-SNR score on the DNS-Challenge dataset, it performs worse on the remaining metrics. Moreover, our proposed method achieves a smaller model size than most of the benchmarks, rendering it more practical for real-world deployment. The processed audio clips can be found at \texttt{https://github.com/Zhangyuewei98/OFIF-Net.git}.

\section{Conclusions}

In this study, we introduce an overlapped-frame information fusion scheme to fully utilize information within inherent algorithmic delay caused by inverse TF transformation. At each frame index, we construct several pseudo future speech frames and fuse them with the current frame to obtain a fused result, which is then fed into the SE network. Additionally, we design a TFCA block to perform feature map recalibration through self-attention-based operations in the time, frequency, and channel dimensions, respectively. Furthermore, we ensure that all parts of the proposed SE system are causal. Experimental results demonstrate the effectiveness of the proposed methods, and the final SE system, namely OFIF-Net, achieves superior performance with a small model size. In the furture, we will attempt to reduce the inherent algorithmic delay of the SE system.

\section*{Acknowledgment}
This work was supported by the special funds of Shenzhen Science and Technology Innovation Commission under Grant No. CJGJZD20220517141400002.

% \clearpage
% \section{References}
\bibliographystyle{IEEEtran}
\bibliography{refs}

\end{document}